# Deep Learning for Fetal Inflammatory Response Diagnosis in the Umbilical Cord

Marina A. Ayad[a], Ramin Nateghi[b], Abhishek Sharma[c], Lawrence Chillrud[a], Tilly Seesillapachai[a], Lee A.D. Cooper[a,c], Jeffery A. Goldstein[a]

[a]*Northwestern University, Department of Pathology, Chicago, IL, USA*
[b]*Northwestern University, Department of Urology, Chicago, IL, USA*
[c]*Chan Zuckerberg Biohub Chicago, IL, USA*

ABSTRACT

*Introduction:* Inflammation of the umbilical cord can be seen as a result of ascending intrauterine infection or other inflammatory stimuli. Acute fetal inflammatory response (FIR) is characterized by infiltration of the umbilical cord by fetal neutrophils, and can be associated with neonatal sepsis or fetal inflammatory response syndrome. Recent advances in deep learning in digital pathology have demonstrated favorable performance across a wide range of clinical tasks, such as diagnosis and prognosis. In this study we classified FIR from whole slide images (WSI).

*Methods:* We digitized 4100 histological slides of umbilical cord stained with hematoxylin and eosin (H&E) and extracted placental diagnoses from the electronic health record. We build models using attention-based whole slide learning models. We compared strategies between features extracted by a model (ConvNeXtXLarge) pretrained on non-medical images (ImageNet), and one pretrained using histopathology images (UNI). We trained multiple iterations of each model and combined them into an ensemble.

*Results:* The predictions from the ensemble of models trained using UNI achieved an overall balanced accuracy of 0.836 on the test dataset. In comparison, the ensembled predictions using ConvNeXtXLarge had a lower balanced accuracy of 0.7209. Heatmaps generated from top accuracy model appropriately highlighted arteritis in cases of FIR 2. In FIR 1, the highest performing model assigned high attention to areas of activated-appearing stroma in Wharton's Jelly. However, other high-performing models assigned attention to umbilical vessels.

*Discussion:* We developed models for diagnosis of FIR from placental histology images, helping reduce interobserver variability among pathologists. Future work may examine the utility of these models for identifying infants at risk of systemic inflammatory response or early onset neonatal sepsis.

# 1. Introduction

## 1.1. The umbilical cord

The umbilical cord acts as a lifeline between the fetus and the placenta, developing as early as the third week and is fully formed by the seventh week[1]. Its main structures include the umbilical vein and two umbilical arteries, all of which are surrounded by a gelatinous extracellular matrix known as the Wharton's Jelly[2]. This matrix is rich in proteoglycans and plays a significant role in supporting the cord's flexibility and integrity through the fetal development. The umbilical arteries carry deoxygenated blood along with the waste products from the fetus to the placenta, where gas and nutrient exchange takes place. The vein then carries the oxygenated, nutrient-rich blood from the placenta back to the fetus.

Given its essential role in fetal development, histopathological examination of the umbilical cord is critical for identifying of any abnormalities or diseases that may have impacted fetal development[3]. Anomalies in the umbilical cord such as single umbilical artery (SUA), thrombosis, and aneurysm have been associated with stillbirth, asphyxia, preeclampsia, gestational diabetes, intrauterine distress and higher risk of congenital malformations[2,4-6].

## 1.2. Fetal Inflammatory Response

Intrauterine inflammation plays a crucial role in adverse neonatal outcomes, more specifically in the context of ascending infections. These infections occur when pathogens reach amniotic cavity from the lower genital tract after rupture of membranes. Microorganisms localized in the decidua of the supracervical region can propagate through the chorioamnionitic membranes. Neutrophils are not normally present in the chorioamnionitic membranes but in response to infection they migrate from the decidua in case of acute chorioamnionitis. Neutrophil infiltration in different regions of the placenta is considered a hallmark of acute inflammatory lesions. A chemotactic gradient drives neutrophil migration into the intervillous space and the chorionic plate, characterized as a maternal inflammatory response (MIR).

In contrast, acute inflammation of the umbilical cord and the chorionic vessels of the chorionic plate is of fetal origin, known as Fetal Inflammatory Response (FIR). The infiltration of neutrophils of fetal origin into the umbilical vessel wall or vessels of the chorionic plate are diagnostic signs of FIR[7]. FIR is linked to elevated levels of Interleukin IL-6 driving elevated concentration of chemokines in the amniotic fluid, resulting in a chemotactic gradient which attracts neutrophils from the umbilical vessel lumens into the Wharton's jelly. The severity of funisitis, an inflammation of the umbilical cord, is correlated with the fetal plasma IL-6 levels, reflecting the intensity of the intraamniotic inflammatory response[8].

FIR is staged into three levels based on the location of neutrophil infiltration. An uninflamed umbilical cord, free of neutrophil infiltration, has no FIR, which we will refer to as FIR 0 (**Fig. 1. A** and **B**). FIR 1, umbilical phlebitis, is characterized by neutrophil infiltration into the muscle layers of the umbilical vein (**Fig. 1.C**). FIR is also diagnosed when there is acute inflammation of the chorionic plate vessels, which we will not further address in this work. FIR 2, umbilical arteritis, is described as neutrophil infiltration into the muscle of the arterial walls (**Fig. 1.E**). FIR

3 involves necrotizing funisitis, inflammation in Wharton's Jelly with circumferential neutrophils or necrosis (**Fig. 1.F** and **G**). Inflammation of Wharton's Jelly without necrosis, "funisitis," may be seen in the context of FIR 1 or 2 when neutrophils migrate past the muscle layers of artery or vein, but does not impact the staging of FIR[9] (**Fig. 1.D**).

Although staging of FIR is important, interobserver variability remains challenging. An overall agreement of 6 examiners showed a Fleiss' κ of 0.73 for stage 2 diagnosis, and an interrater correlation coefficient of 0.68 for stages 1 to 3[10]. Another study compared general practice surgical pathologists with an expert placental pathologist and showed a Cohen's kappa of 0.49 for any diagnosis of FIR (FIR0 vs FIR1,2,3)[11]. The development of diagnostic tools could mitigate the scarcity of expert placental pathologists and improve diagnostic consistency.

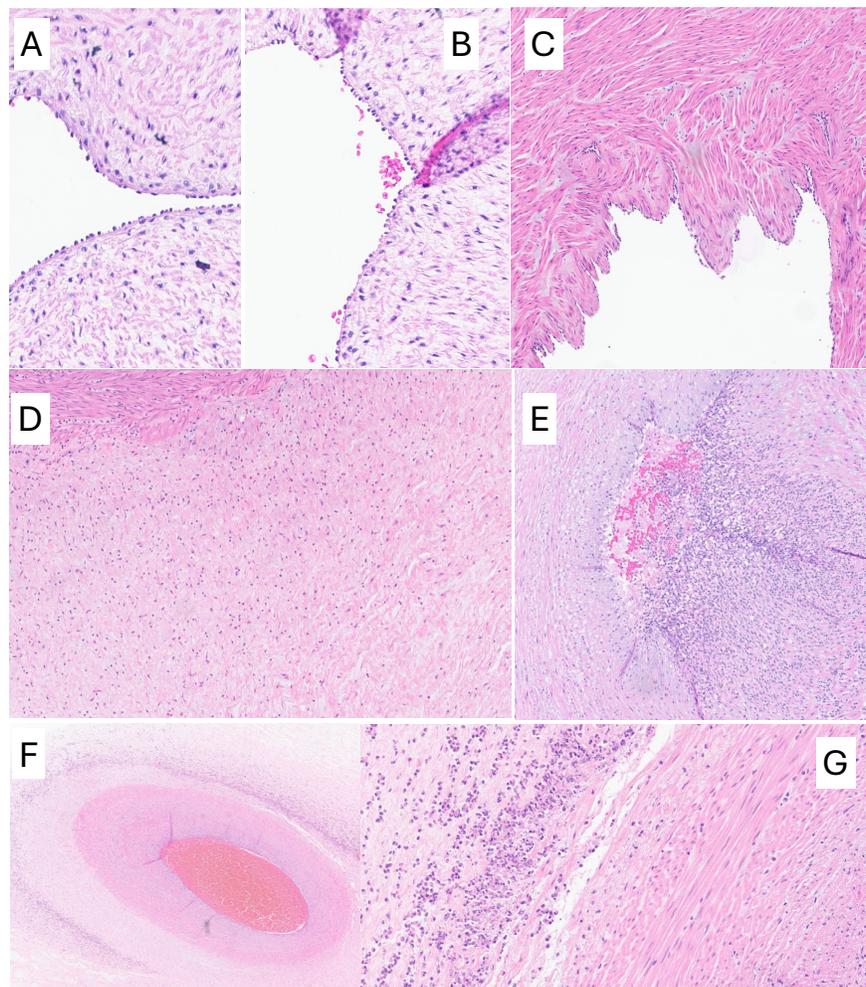

**Fig. 1.** FIR staging in the umbilical cord. A. Normal umbilical cord, no inflammation in umbilical vein (20×). B: No Inflammation in umbilical artery. C: FIR 1: Acute umbilical phlebitis with amniotropic migration of fetal neutrophils in the muscle layer of umbilical vein (10×). D. Funisitis showing inflammation beyond the vessels and neutrophils present in Wharton's jelly(10×). E: FIR 2: Umbilical arteritis with concomitant phlebitis (10×). F: FIR 3: Necrotizing funisitis showing concentric(2.5×), perivascular distribution of necrotizing neutrophils around vessels. G: FIR 3: Necrotic neutrophils in Wharton's jelly(20×)

*1.3. Deep learning in digital pathology diagnostics*

The field of AI in diagnosis and prognosis of pathology images has shown rapid advances [12-14]. With the growing demand of precision medicine and the increase in treatment options, AI allows the quantitative extraction of information from whole slide images (WSI) across various medical specialties. AI algorithms assist in tasks such as biomarker detection[15-18], prognosis[19-21], and diagnosis[22-24]. One of the major promising advantages of AI in digital pathology is its potential to reduce interobserver variability which can improve prognostic and treatment outcomes especially in regions with limited access to specialized pathologists and medical resources[25-29].

A major computational challenge in digital pathology is the large size of the scanned WSIs which averages 120,000 × 80,000 pixels significantly larger than traditional computational analysis tasks that are typically 256 × 256-pixel images[30,31]. Attention-based multiple instance learning (MIL) approaches have been shown to yield highly accurate models for generating whole-slide diagnoses. [32-34]. This mechanism allows models to be trained using patient or WSI-level labels instead of patch-level labels minimizing the annotation burden on pathologists and easing future clinical implementation. A useful byproduct of attention-based approaches is that they assign an attention value to each portion of the image, based on the degree to which each portion of the image contributes to the model output. Despite the risk of over-interpretation, examination of these models can aid in validating or troubleshooting model performance[35].

In this study, we developed ML models to classify the FIR stages using WSI of umbilical cord. The attention maps generated by these models provide a visualization of their ability to identify critical regions in the umbilical cord, such as umbilical vessels and differentiate between the arteries and the vein, to diagnose FIR.

## 2. Methods

*2.1. Patients and diagnosis*

We identified patients who underwent delivery and examination of the placenta and umbilical cord in our institution between years 2011 and 2023. Slides were digitized on a Leica GT450 scanner with a 40× objective magnification (0.263 μm per pixel). The corresponding placental pathology reports were obtained from the institutional Electronic Data Warehouse (EDW). Natural Language Processing (NLP) was used to parse the reports[36]. Our inclusion criteria consist of singleton deliveries 12-42 week gestation. Twins or higher multiples, cases with no umbilical cords, and surgical terminations were excluded from this study. Patients and diagnostic data were stored in REDcap[37]. The study was approved by Northwestern institutional review board (STU00214052).

## 2.2. Model overview

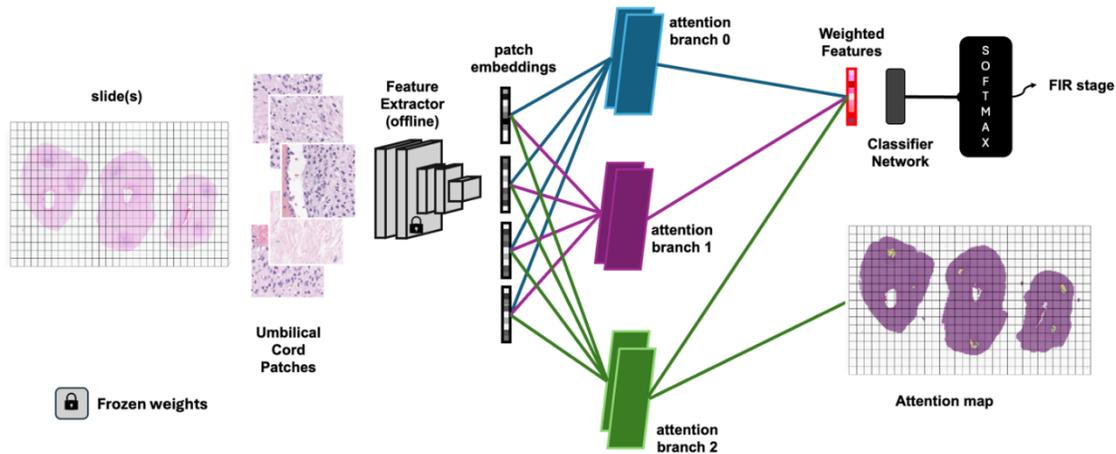

**Fig. 2.** Model Overview Diagram: Tissue regions in whole slide images are divided into patches. Each patch passes through a feature extraction network to extract patch embeddings. The extracted features pass through an attention branch for each FIR stage. Two outputs are produced. First, the attention subnetwork generates attention values for each feature vector and is then used to aggregate weighted features, which passes through the classifier network for FIR prediction. Second, the attention values for each patch for the predicted class is used to plot an attention map. The sample shown in the figure is FIR 2, showing high attention at the arteries.

Our model is divided into two sections, the first is a feature extraction which transforms the slides into embedding in the latent space to be used for down streaming tasks. The second part is the classification using weakly supervised whole-slide learning. This model provides both the final classification and attention score for each patch in the slide. Due to the rarity of FIR 3 slides, we combined FIR 3 with FIR 2, resulting in a 3-level classification (FIR 0, FIR 1, FIR 2,3).

### 2.2.1. Feature extraction encoders

One slide per patient was used. Tissue regions were segmented using Otsu's method on low-power slide images[38]. Segmented tissue regions at 20× magnification was divided into 224×244 pixel patches, which resulted in 208-33650 patches per slide and passed for feature extraction. Conventional approaches in pathology machine learning have performed feature extraction using models pretrained on the ImageNet collection of everyday objects[39]. More recently, several groups have released foundation models, that have been trained on very large datasets of histology images from diverse tissues [40-44]. We compared feature extraction from the ConvNeXtXLarge model[45] trained on ImageNet[39] and the pathology foundation model UNI[46]NI was trained using a self-supervised learning approach on 100,000 slides across multiple tissue types and consistently ranks among the top performance foundational model across different tissue types and downstream tasks[47,48]. Significantly for our application, the UNI dataset does not include placenta.

To compare the feature embedding we applied a zero-shot classification of regions in the umbilical cord. We began by extracting features using ConvNextXLarge model pretrained by ImageNet from 20 slides of validation dataset. After normalizing the features, we applied a principal component analysis (PCA) for dimensionality reduction. We then performed a K-means clustering with 5 clusters. From each cluster, we selected 10 patches per slide, resulting in a total of 1000 patches. A perinatal pathologist divided the patches into six groups: amnion, blood, vessel wall, Wharton's jelly, non-umbilical cord tissue, and debris/marker. We extracted the features of these annotated patches and visualized them using t-Distributed Stochastic Neighbor Embedding (t-SNE). To evaluate the classification performance, we applied a k-Nearest Neighbor (KNN) Classifier and calculated the balanced accuracies.

### 2.2.2. Attention-based multiple instance learning

We used attention-based MIL approach for classification and attention map evaluation (**Fig. 2**). In this framework, the WSI is divided into smaller patches - 224x224 pixels in our case. The WSI is treated as a collection of multiple patches, each of which may or may not be informative for the diagnosis. In the first step, patches are embedded using the feature extractor – either UNI or ConvNeXtXLarge. The patch embeddings, represented as a list of numbers describing the features, are passed through a batch normalization layer followed by a dimensionality reduction layer using a fully connected layers with 512 outputs and a rectified linear unit (ReLU) activation. The activation is diverged into two branches with a fully connected dense layer of 256 output, one using a tanh activation and the other using a sigmoid activation. The tanh and sigmoid are two non-linear functions that gives the model the ability to capture different aspects of the input. An element-wise product on the outputs of both branches is evaluated and used as an input for the multi-class attention section. We used 3 branches – 1 each for the diagnoses of FIR 0, FIR 1, and FIR 2,3 - with a fully connected layer of 64 output followed by a ReLU activation. A linear activation is applied, and the output of the linear layer is passed through a sigmoid activation function. The attention mechanism assigns weights to different features based on their importance. A weighted average pooling is applied to the outputs of all three branches by applying an element-wise multiplication and normalization. This is then saved for the attention maps and passed for the classification task. The classification network is a linear activation layer followed by a softmax activation.

### 2.3. Hyperparameter Optimization Experiments

A fundamental challenge in deep learning design, is hyperparameter optimization. Searching the optimal combination of optimization algorithm, optimization parameters, number of epochs, learning rate, and use of exponential moving average and momentum is not straightforward, and trial and error is inefficient and time consuming. We used population-based training (PBT) to train a group of models concurrently, allowing rapid recognition of the optimal settings[49]. Initially, neural networks are trained in parallel with random hyperparameters. During training of the population of networks, parameters of successful networks are copied ("exploited") and then modified ("exploring"). This way PBT can dedicate more training time and usage of computational resources in more promising models. We used the Ray-tune package for PBT with the target metric of multiclass balanced accuracy, minimum 15 epochs, and hinge loss function[50]. Class weights were used to counter the effect of unbalanced classes.

Our hyperparameter search space included optimization algorithms, such as RMSprop, SGD, Adam, Adagrad along with a range of learning rates and learning rate decay values for each optimizer. Specifically for the Adam optimizer, the search space included a range of $\beta_1$ and $\beta_2$ values. Additionally, it included the optional usage of EMA and specified a momentum value to control the influence of prior iterations when updating model weights. Higher momentum values will weigh the prior iterations more. A momentum range is provided to determine the frequency of overwriting the model weights with the calculated averages.

*2.4.  Model Ensembles*

Population-based training results in multiple models that not only have different performance but may capture different aspects of the underlying biology[51-54]. By averaging the results of models with different strengths, higher performance may be seen. For ConvNextXLarge and UNI, we trained 900 models each. To determine how many models to include in our ensemble, we empirically tested for the group with the highest accuracy. Ensemble predictions were performed by weighting model outputs based on the model's AUROC.

*2.5.  Attention Maps*

Each branch of the model – FIR0, FIR1, FIR2,3 – assigns an attention value to each patch. These values can be plotted to show the relative contribution of different parts of the umbilical cord to the final result.

## 3. Results

We developed a dataset of 4100 slides, which included 3337 umbilical cords with no inflammation, FIR 0, 480 as FIR 1, 252 as FIR 2, and 31 as FIR 3. Due to the low numbers of FIR3, we combined those cases with FIR2, resulting in three-classes. The dataset was stratified and split into an 80% training, 10% validation, 10% testing sets. **Table 1** displays the dataset distribution. A one-way ANOVA was calculated to assess difference in maternal and gestational age across the three FIR classes, revealing significant difference ($p < 0.05$). Additionally, the counts of parous individuals showed a significant difference, as calculate be a chi-square test ($\chi^2$). In contrast, fetal sex showed no significant difference across FIR classes.

**Table 1**
Dataset Demographics:

|  | FIR 0, n = 3337 | FIR 1, n = 480 | FIR 2,3, n = 283 | p-value |
|---|---|---|---|---|
| **Maternal Age (years)** | 33(15-55) | 33(17-49) | 32(17-50) | 0.001(ANOVA) |
| **Nulliparous** | 1836 (55%) | 327 (68%) | 194 (69%) | $3.47 \times 10^{-7}(\chi^2$ test) |
| **Gestational Age (days)** | 37.9(13-42) | 39.3(19.3-42.4) | 37.3(15.1-42) | $1.98 \times 10^{-25}$ (ANOVA) |
| **Fetal Sex** | Female: 1798 (54%)<br>Male: 1539 (46%) | 222 (46%)<br>258 (54%) | 133 (47%)<br>150 (53%) | $0.0397(\chi^2$ test) |

**Values are median (IQR) or count (percent) as appropriate**

## 3.1. Model Performance

For classification, we implemented MIL using both the ConvNeXtXLarge pretrained with ImageNet and UNI features. To evaluate the impact of different embeddings, we utilized the same hyperparameter optimization pipeline for both models. The top performing UNI model achieved a balanced accuracy of 0.7993 (**Table 2**), higher than the best result from ConvNeXtXLarge embeddings, balanced accuracy of 0.72 (**Table 3**).

**Table 2**
Summary of Results with ConvNeXtXLarge embeddings pretrained with ImageNet including AUC, Balanced Accuracy, Sensitivity, Specificity for top models with balanced accuracies greater than 0.7. Weighted average ensembling results in higher performance.

| Model Rank | AUC | Balanced Accuracy | Sensitivity | Specificity |
|---|---|---|---|---|
| 1 | 0.828 | 0.723 | 0.723 | 0.84 |
| 2 | 0.9 | 0.706 | 0.706 | 0.855 |
| 3 | 0.787 | 0.701 | 0.701 | 0.837 |
| 4 | 0.872 | 0.7 | 0.7 | 0.837 |
| **Ensemble** | **0.901** | **0.721** | **0.721** | **0.853** |

**Table 3**
Summary of Results with UNI embeddings including AUC, Balanced Accuracy, Sensitivity, Specificity for top models with balanced accuracies greater than 0.77. Weighted average ensembling results in higher performance.

| Model Rank | AUC | Balanced Accuracy | Sensitivity | Specificity |
|---|---|---|---|---|
| 1 | 0.928 | 0.799 | 0.799 | 0.914 |
| 2 | 0.906 | 0.795 | 0.795 | 0.904 |
| 3 | 0.903 | 0.790 | 0.79 | 0.897 |
| 4 | 0.923 | 0.779 | 0.779 | 0.897 |
| 5 | 0.914 | 0.779 | 0.779 | 0.896 |
| 6 | 0.888 | 0.774 | 0.774 | 0.898 |
| **Ensemble** | **0.923** | **0.836** | **0.836** | **0.92** |

## 3.2. Ensembling

The ensembling the top 4 ConvNeXtXLarge trained models resulted in a balanced accuracy of 0.72. In comparison, the ensemble of UNI models achieved a balanced accuracy of 0.836. This was markedly higher than the individual models, suggesting that, despite similar accuracy values, the models have complementing strengths. The confusion matrices of the ensembled models of both embeddings are shown in **Fig. 3**.

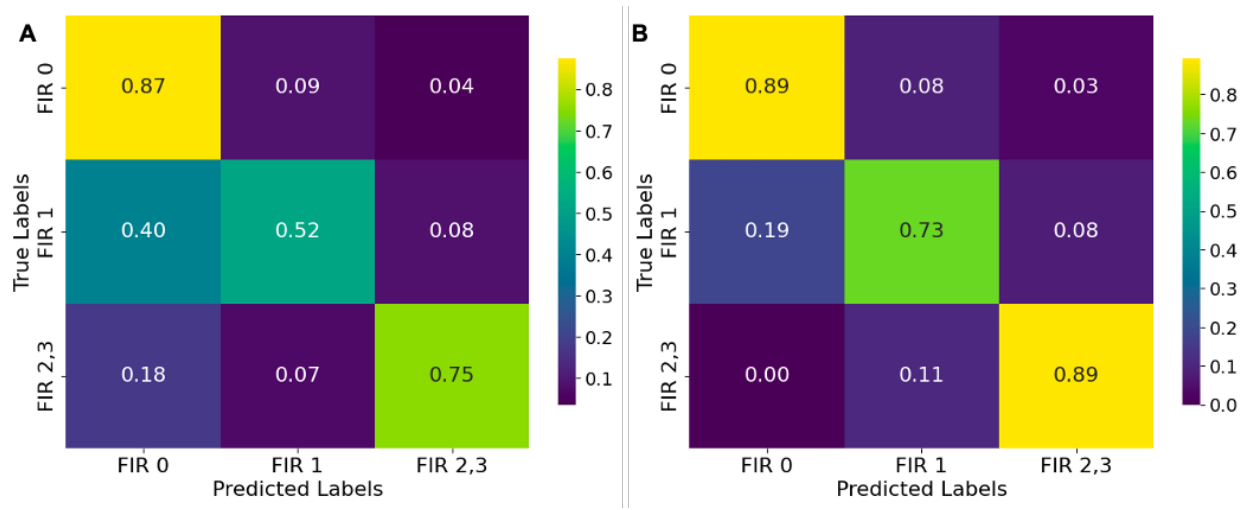

**Fig. 3.** Confusion Matrices for A) ConvNeXtXLarge Model Ensembling Predictions, B) UNI Model Ensembling Predictions. Values are the proportion of FIR 0, FIR 1, and FIR 2,3 classes.

## 3.3. Attention Maps and explainability

Interpretation of AI decision-making processes may be challenging. Our models assign an attention value to each patch in the WSI, which reflects the contribution of that patch to the model's output. Extracting the attention values can highlight the important patches – but not why they are important or how the model is using them. Within the ensembles, each models generates its own attention maps. We began by analyzing the attention values of the top performing member of the UNI ensemble (Representative images: **Fig. 4**, Full images: **Supplementary Fig. 1-3**), showing samples that were correctly predicted by that model. In FIR 0, representing uninflamed umbilical cord, the model's attention was low and diffusely distributed across the umbilical cord. The top two attention patches were unremarkable Wharton's jelly, and umbilical cord surface, respectively, which reflects the nonspecific focus of this attention. In FIR 1, attention maps show the highest focus within Wharton's Jelly. The umbilical veins, which are critical for the diagnosis of FIR 1, show low attention. For a case of FIR 2, the model focuses on portions of the umbilical arterial wall with neutrophilic inflammation – reflecting the portions of the image that a pathologist would use to make the diagnosis of arteritis.

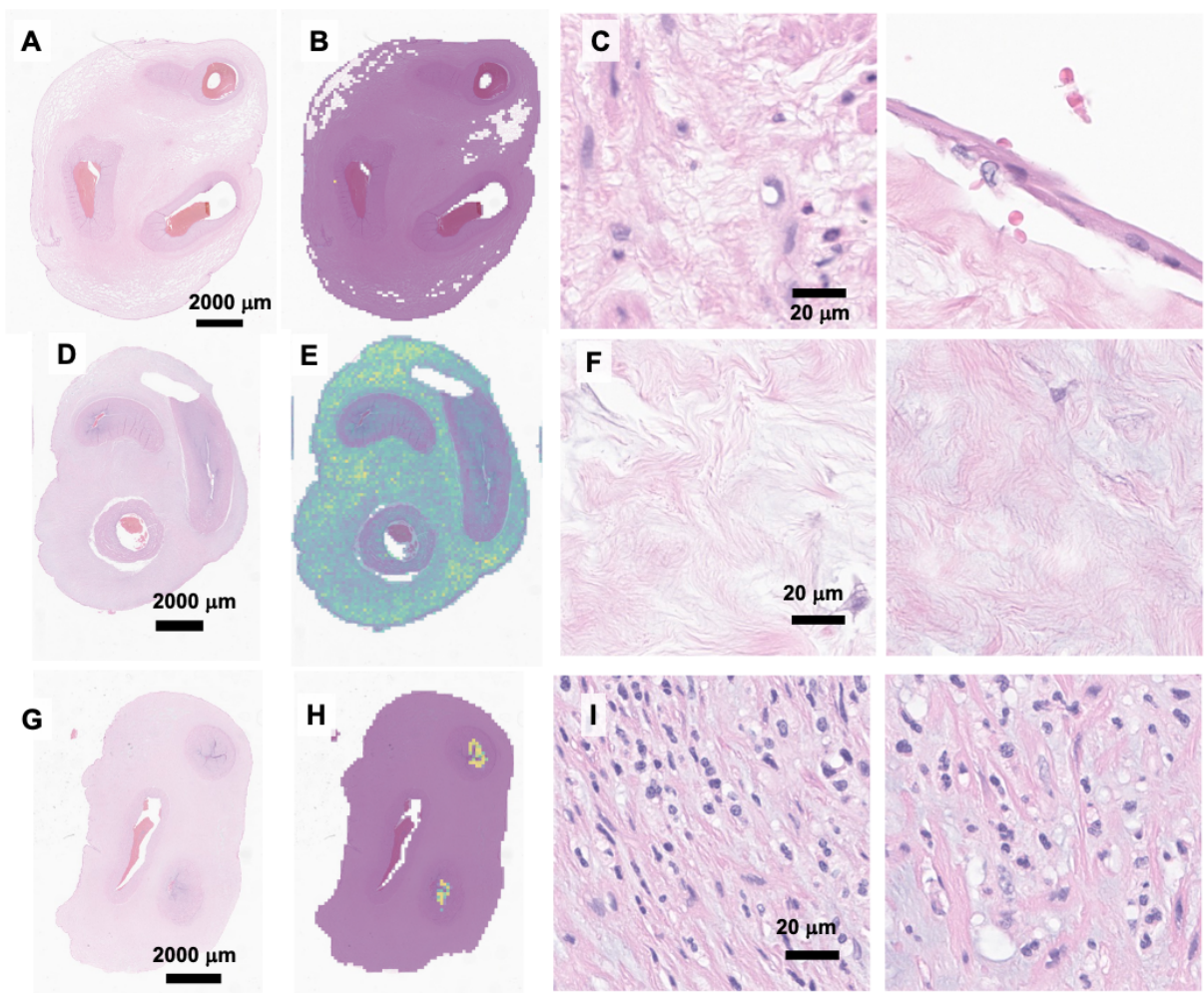

**Fig. 4.** Attention Maps of FIR predictions. A) FIR 0 umbilical cord sample (low attention: purple), B) The attention map of FIR 0 shows a uniform distribution of attention across the tissue, C) Top 2 high-attention patches are drawn from different, non-specific regions, in the umbilical cord tissue. D) FIR 1 sample, E) The attention map for FIR 1 highlights regions across the stroma of the Wharton's Jelly, F) Top 2 high-attention patches, show a mixture of normal and activated-appearing stroma within the Wharton's jelly. (high attention: yellow) G) FIR 2 sample with umbilical arteritis, H) The attention map shows concentrated focus in the arteries, I) Top 2 high attention patches show inflammation in the umbilical arterial walls

### 3.4. How does the UNI ensemble recognize FIR 1?

The UNI ensemble has good performance in diagnosing FIR 1, despite the difficult-to-justify attention map produced by the top performing model. We therefore examined the output and attention maps of the other 5 members of the ensemble. 5/6 models in the ensemble gave the correct prediction, while the #4 ranked model gave an incorrect prediction. Examining the attention maps showed significant variability (**Fig. 5**). The top performing model as well as the 3rd and 4th ranked models showed attention in Wharton's Jelly. The 2nd and 5th highest ranked models showed moderate-high attention in the vessels. Unlike FIR 2, where the models attend to the arteries specifically, the FIR 1 models that attend to vessels do so equally between the arteries and the vein. The 6th ranked model showed diffuse low attention across the cord, differing from the other members of the ensemble. We examined several cases of FIR 1, which showed similar patterns of attention (Not Shown). In contrast to FIR 1, The different members of the ensemble showed very similar attention maps for FIR 0 and FIR 2,3 was more consistent. All six models showed low, diffuse attention for FIR 0, and high attention on arteries for FIR 2,3 (**Supplementary Fig. 4** and **5**, respectively).

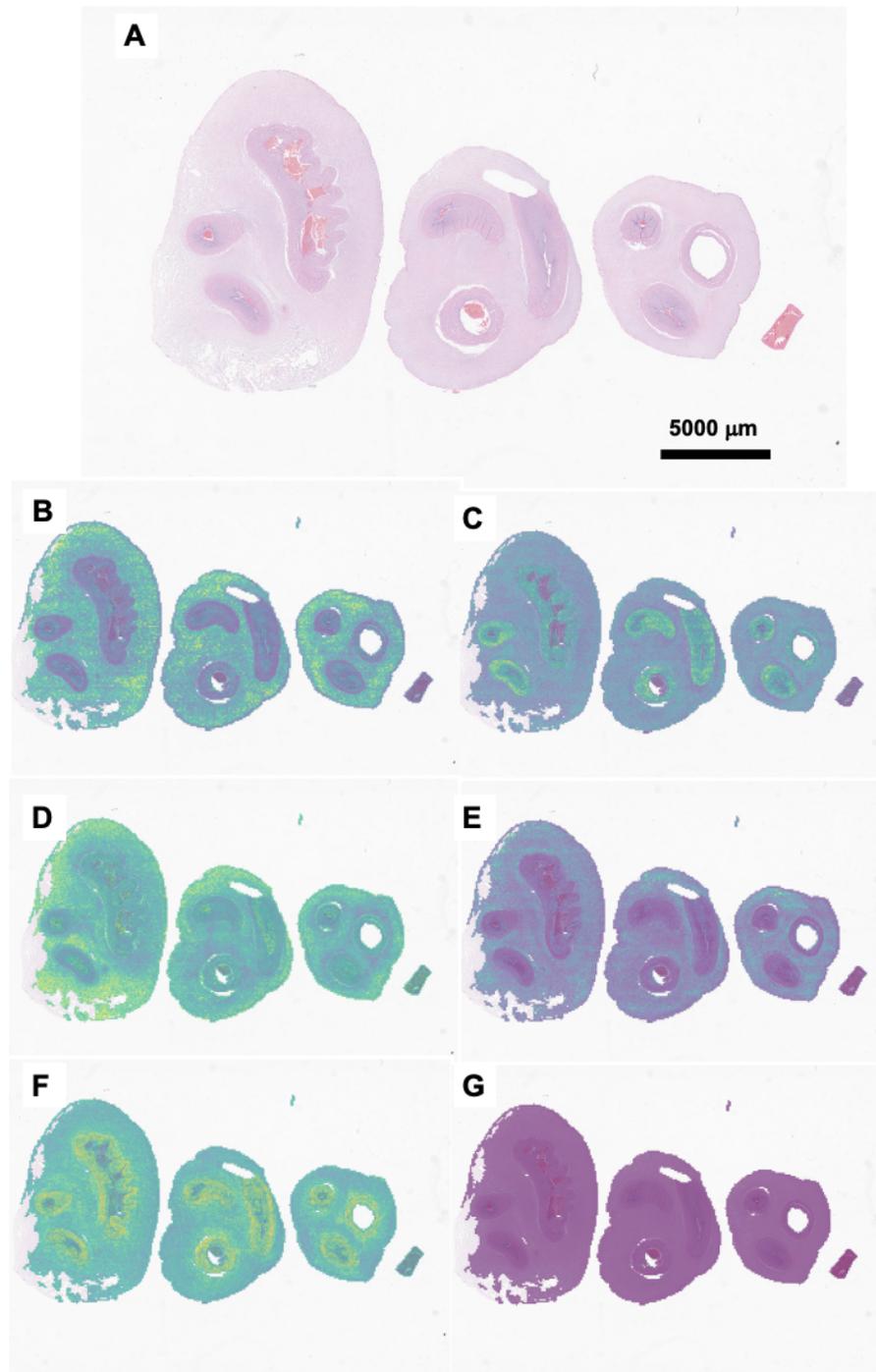

**Fig. 5.** FIR 1 Sample. A: The original histology image. B-G: Attention maps for models ranked from the highest balanced accuracy model to the lowest, as included in the ensemble. Models in B, D, and E shows attention in stroma of Wharton's Jelly. Models C, and F shows attention in around the vessels.

## 3.5. Feature Embeddings Analysis

To further examine why UNI shows higher performance in classification of FIR, we compared the ability of feature extractors to classify patches from different regions of the umbilical cord. We used a clustering approach, in which the class of a tissue region is inferred based how its image features, as extracted by a feature extractor, cluster. We examined the clustering using t-SNE (**Fig. 6**). t-SNE projects the 1024-dimensional feature vectors of ConvNeXtXLarge and UNI into two dimensional space and places points with similar feature vectors closer together. The t-SNE for ConvNeXtXLarge show some separation of points by class, but with significant intermixture. In contests, plots demonstrate tight clusters for the features extracted by UNI. The balanced accuracy of KNN for ConvNeXtXLarge was 0.6522. For UNI it was 0.9681. These results highlight how foundational models pretrained using histopathology images result in a better classification, even though UNI was not pretrained on placental tissue.

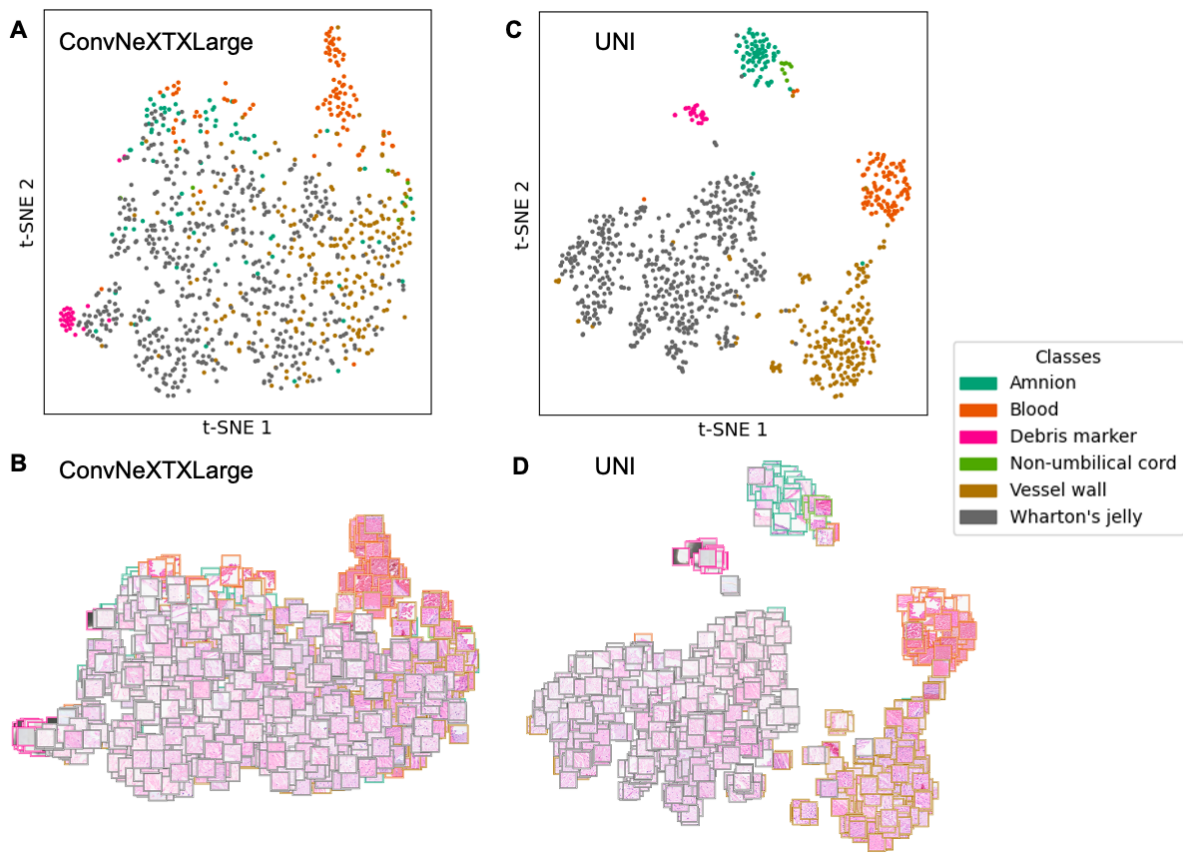

**Fig. 6.** Feature Embeddings. A: t-SNE plot for ConvNeXtXLarge features with color-coded class labels and B: the extracted patches. C: t-SNE plot of UNI features, demonstrating clearer cluster separation and D: extracted patches.

## 4. Discussion

We developed machine learning-based classifiers that identify FIR in the umbilical cord. We show that grouping models together in ensembles improves performance and that differences in attention mechanism may allow models to compensate for one another. We show that the use of a foundation model, trained on pathology images, improves performance in this diagnosis, even though the model was not trained using placenta. Finally, we suggest that this difference may be due to better distinction between different classes of tissue at baseline.

Our method does not require manual annotation of the specific regions such as the arteries nor the vessel in the umbilical cord. We process the WSI with the FIR stage diagnosis as the labels. Our slides consist of two or three slices of the umbilical cord with different morphologies. One computational challenge is that inflammation is typically focal – localized to only a single or few vessels. An ideal model would assign high attention to these areas, particularly those that are informative for a particular diagnosis. This behavior is seen in FIR 2,3 models. A perquisite for this behavior is successfully distinguishing umbilical arterial and venous tissue, which is not always straightforward.

In contrast, none of the high-performing models attend to the vein specifically in FIR 1. Three models attended to Wharton's Jelly, two attended to all vessels, and one showed diffuse low attention. We are confronted by two possibilities. Artificial intelligence has shown the potential to increase performance by highlighting regions in the tissue that have been overlooked by pathologists[12,21,55]. For example, a deep learning model identified stroma in non-cancerous area of prostate tissue to be a prognostic factor in the histology of prostate cancer[55]. Another model showed that breast cancer prognosis worsens when cancer-associated fibroblasts are located closer the invasive edge of the tumor[56].

Alternatively, and perhaps more likely, is that this is an example of "shortcut learning" - the model has failed to encode the underlying biological process and is relying on correlations that are not directly related to the diagnosis.[57] Inflammation in Wharton's Jelly or the umbilical arteries is almost invariably accompanied by phlebitis, so the apparent strategies of focusing on Wharton's Jelly and any vessel can work. Even low diffuse attention could work in some circumstances – finding a neutrophil anywhere on the slide is correlated to the diagnosis of phlebitis. Combining multiple models in an ensemble resulted in improved performance, however classification of FIR 1 was still a weak point for the ensemble, perhaps due to the apparent reliance on shortcuts.

## 5. Conclusion

In conclusion, we developed an attention-based multiple instance learning model, a weakly supervised method, to predict FIR stage from the WSI of umbilical cords. Using population-based training we generated an ensemble of models to enhance prediction accuracy. Additionally, we generated attention maps corresponding to the predicted FIR stages providing insights the regions that contributed to the model's decision-making mechanism.

# References

1. Heil JR BB. Embryology, Umbilical Cord. *StatPearls* 2023;doi:https://www.ncbi.nlm.nih.gov/books/NBK557490/
2. Debebe SK, Cahill LS, Kingdom JC, et al. Wharton's jelly area and its association with placental morphometry and pathology. *Placenta*. May 2020;94:34-38. doi:10.1016/j.placenta.2020.03.008
3. Redline RW. Inflammatory responses in the placenta and umbilical cord. *Semin Fetal Neonatal Med*. Oct 2006;11(5):296-301. doi:10.1016/j.siny.2006.02.011
4. Dubetskyi BI, Makarchuk OM, Zhurakivska OY, et al. Pregnancy and umbilical cord pathology: structural and functional parameters of the umbilical cord. *J Med Life*. Aug 2023;16(8):1282-1291. doi:10.25122/jml-2023-0025
5. Siargkas A, Giouleka S, Tsakiridis I, et al. Prenatal Diagnosis of Isolated Single Umbilical Artery: Incidence, Risk Factors and Impact on Pregnancy Outcomes. *Medicina (Kaunas)*. Jun 3 2023;59(6)doi:10.3390/medicina59061080
6. Iwagaki S, Takahashi Y, Chiaki R, et al. Umbilical cord length affects the efficacy of amnioinfusion for repetitive variable deceleration during labor. *J Matern Fetal Neonatal Med*. Jan 2022;35(1):86-90. doi:10.1080/14767058.2020.1712703
7. Khong TY, Mooney EE, Ariel I, et al. Sampling and Definitions of Placental Lesions: Amsterdam Placental Workshop Group Consensus Statement. *Arch Pathol Lab Med*. Jul 2016;140(7):698-713. doi:10.5858/arpa.2015-0225-CC
8. Kim CJ, Yoon BH, Romero R, et al. Umbilical arteritis and phlebitis mark different stages of the fetal inflammatory response. *American Journal of Obstetrics and Gynecology*. 2001/08/01/ 2001;185(2):496-500. doi:https://doi.org/10.1067/mob.2001.116689
9. Goldstein JA, Gallagher K, Beck C, Kumar R, Gernand AD. Maternal-Fetal Inflammation in the Placenta and the Developmental Origins of Health and Disease. *Front Immunol*. 2020;11:531543. doi:10.3389/fimmu.2020.531543
10. Redline RW, Vik T, Heerema-McKenney A, et al. Interobserver Reliability for Identifying Specific Patterns of Placental Injury as Defined by the Amsterdam Classification. *Arch Pathol Lab Med*. Mar 1 2022;146(3):372-378. doi:10.5858/arpa.2020-0753-OA
11. Ernst LM, Basic E, Freedman AA, Price E, Suresh S. Comparison of Placental Pathology Reports From Spontaneous Preterm Births Finalized by General Surgical Pathologists Versus Perinatal Pathologist: A Call to Action. *Am J Surg Pathol*. Oct 1 2023;47(10):1116-1121. doi:10.1097/pas.0000000000002111
12. Baxi V, Edwards R, Montalto M, Saha S. Digital pathology and artificial intelligence in translational medicine and clinical practice. *Modern Pathology*. 2022/01/01 2022;35(1):23-32. doi:10.1038/s41379-021-00919-2
13. McGenity C, Clarke EL, Jennings C, et al. Artificial intelligence in digital pathology: a systematic review and meta-analysis of diagnostic test accuracy. *npj Digital Medicine*. 2024/05/04 2024;7(1):114. doi:10.1038/s41746-024-01106-8
14. Shafi S, Parwani AV. Artificial intelligence in diagnostic pathology. *Diagnostic Pathology*. 2023/10/03 2023;18(1):109. doi:10.1186/s13000-023-01375-z

15. Lancellotti C, Cancian P, Savevski V, et al. Artificial Intelligence & Tissue Biomarkers: Advantages, Risks and Perspectives for Pathology. *Cells*. Apr 2 2021;10(4)doi:10.3390/cells10040787
16. Li Z, Bui MM, Pantanowitz L. Clinical tissue biomarker digital image analysis: A review of current applications. *Human Pathology Reports*. 2022/06/01/ 2022;28:300633. doi:https://doi.org/10.1016/j.hpr.2022.300633
17. Chang X, Wang J, Zhang G, et al. Predicting colorectal cancer microsatellite instability with a self-attention-enabled convolutional neural network. *Cell Rep Med*. Feb 21 2023;4(2):100914. doi:10.1016/j.xcrm.2022.100914
18. Jiang S, Zanazzi GJ, Hassanpour S. Predicting prognosis and IDH mutation status for patients with lower-grade gliomas using whole slide images. *Scientific Reports*. 2021/08/19 2021;11(1):16849. doi:10.1038/s41598-021-95948-x
19. Xu Z, Lim S, Shin H-K, et al. Risk-aware survival time prediction from whole slide pathological images. *Scientific Reports*. 2022/12/19 2022;12(1):21948. doi:10.1038/s41598-022-26096-z
20. Liu H, Kurc T. Deep learning for survival analysis in breast cancer with whole slide image data. *Bioinformatics*. 2022;38(14):3629-3637. doi:10.1093/bioinformatics/btac381
21. Amgad M, Hodge J, Elsebaie M, et al. A population-level computational histologic signature for invasive breast cancer prognosis. *Res Sq*. May 26 2023;doi:10.21203/rs.3.rs-2947001/v1
22. Goldstein JA, Nateghi R, Irmakci I, Cooper LAD. Machine learning classification of placental villous infarction, perivillous fibrin deposition, and intervillous thrombus. *Placenta*. Apr 2023;135:43-50. doi:10.1016/j.placenta.2023.03.003
23. Lee M. Recent Advancements in Deep Learning Using Whole Slide Imaging for Cancer Prognosis. *Bioengineering (Basel)*. Jul 28 2023;10(8)doi:10.3390/bioengineering10080897
24. Al-Thelaya K, Gilal NU, Alzubaidi M, et al. Applications of discriminative and deep learning feature extraction methods for whole slide image analysis: A survey. *J Pathol Inform*. 2023;14:100335. doi:10.1016/j.jpi.2023.100335
25. Borrelli P, Ulen J, Enqvist O, Edenbrandt L, Tragardh E. AI tool decreases inter-observer variability in the analysis of PSMA-PET/CT. *Journal of Nuclear Medicine*. 2021;62(supplement 1):1006-1006.
26. Tizhoosh HR, Diamandis P, Campbell CJV, et al. Searching Images for Consensus: Can AI Remove Observer Variability in Pathology? *The American Journal of Pathology*. 2021/10/01/ 2021;191(10):1702-1708. doi:https://doi.org/10.1016/j.ajpath.2021.01.015
27. Goldenberg RL, McClure EM, Bhutta ZA, et al. Stillbirths: the vision for 2020. *The Lancet*. 2011/05/21/ 2011;377(9779):1798-1805. doi:https://doi.org/10.1016/S0140-6736(10)62235-0
28. Luchini C, Pantanowitz L, Adsay V, et al. Ki-67 assessment of pancreatic neuroendocrine neoplasms: Systematic review and meta-analysis of manual vs. digital pathology scoring. *Modern Pathology*. 2022/06/01/ 2022;35(6):712-720. doi:https://doi.org/10.1038/s41379-022-01055-1
29. López-Pérez M, Amgad M, Morales-Álvarez P, et al. Learning from crowds in digital pathology using scalable variational Gaussian processes. *Scientific Reports*. 2021/06/02 2021;11(1):11612. doi:10.1038/s41598-021-90821-3
30. Madabhushi A, Lee G. Image analysis and machine learning in digital pathology: Challenges and opportunities. *Med Image Anal*. Oct 2016;33:170-175. doi:10.1016/j.media.2016.06.037

31. Tizhoosh HR, Pantanowitz L. Artificial Intelligence and Digital Pathology: Challenges and Opportunities. *J Pathol Inform*. 2018;9:38. doi:10.4103/jpi.jpi_53_18
32. Rakha EA, El-Sayed ME, Lee AHS, et al. Prognostic Significance of Nottingham Histologic Grade in Invasive Breast Carcinoma. *Journal of Clinical Oncology*. 2008/07/01 26(19):3153-3158. doi:10.1200/JCO.2007.15.5986
33. Teramoto A, Kiriyama Y, Tsukamoto T, et al. Weakly supervised learning for classification of lung cytological images using attention-based multiple instance learning. *Scientific Reports*. 2021/10/13 2021;11(1):20317. doi:10.1038/s41598-021-99246-4
34. Lu MY, Williamson DFK, Chen TY, Chen RJ, Barbieri M, Mahmood F. Data-efficient and weakly supervised computational pathology on whole-slide images. *Nature Biomedical Engineering*. 2021/06/01 2021;5(6):555-570. doi:10.1038/s41551-020-00682-w
35. Rudin C. Stop explaining black box machine learning models for high stakes decisions and use interpretable models instead. *Nature Machine Intelligence*. 2019/05/01 2019;1(5):206-215. doi:10.1038/s42256-019-0048-x
36. Shanes ED, Mithal LB, Otero S, Azad HA, Miller ES, Goldstein JA. Placental Pathology in COVID-19. *Am J Clin Pathol*. Jun 8 2020;154(1):23-32. doi:10.1093/ajcp/aqaa089
37. Harris PA, Taylor R, Minor BL, et al. The REDCap consortium: Building an international community of software platform partners. *J Biomed Inform*. Jul 2019;95:103208. doi:10.1016/j.jbi.2019.103208
38. Otsu N. A Threshold Selection Method from Gray-Level Histograms. *IEEE Transactions on Systems, Man, and Cybernetics*. 1979;9(1):62-66. doi:10.1109/TSMC.1979.4310076
39. Deng J, Dong W, Socher R, Li LJ, Kai L, Li F-F. ImageNet: A large-scale hierarchical image database. 2009:248-255.
40. Moor M, Banerjee O, Abad ZSH, et al. Foundation models for generalist medical artificial intelligence. *Nature*. 2023/04/01 2023;616(7956):259-265. doi:10.1038/s41586-023-05881-4
41. Lu MY, Chen B, Williamson DFK, et al. A visual-language foundation model for computational pathology. *Nature Medicine*. 2024/03/01 2024;30(3):863-874. doi:10.1038/s41591-024-02856-4
42. Xu H, Usuyama N, Bagga J, et al. A whole-slide foundation model for digital pathology from real-world data. *Nature*. 2024/06/01 2024;630(8015):181-188. doi:10.1038/s41586-024-07441-w
43. Vorontsov E, Bozkurt A, Casson A, et al. A foundation model for clinical-grade computational pathology and rare cancers detection. *Nature Medicine*. 2024/07/22 2024;doi:10.1038/s41591-024-03141-0
44. Huang Z, Bianchi F, Yuksekgonul M, Montine TJ, Zou J. A visual–language foundation model for pathology image analysis using medical Twitter. *Nature Medicine*. 2023/09/01 2023;29(9):2307-2316. doi:10.1038/s41591-023-02504-3
45. Liu Z, Mao H, Wu CY, Feichtenhofer C, Darrell T, Xie S. A ConvNet for the 2020s. 2022:11966-11976.
46. Chen RJ, Ding T, Lu MY, et al. Towards a general-purpose foundation model for computational pathology. *Nature Medicine*. 2024/03/01 2024;30(3):850-862. doi:10.1038/s41591-024-02857-3
47. Chad Vanderbilt and Thomas J. Fuchs SCaGCaAEaASaJZaADPaAJSaK-lHaJHa. Benchmarking Embedding Aggregation Methods in Computational Pathology: A Clinical Data Perspective. *arXiv/240707841*. 2024;

48. Smith S, Brock A. et al. ConvNets Match Vision Transformers at Scale. *arXiv/231016764*. 2023;
49. Jaderberg M, Dalibard V. et al. Population Based Training of Neural Networks. <https://arxivorg/abs/171109846>. 2017;
50. Liaw RaL, Eric and Nishihara, Robert and Moritz, Philipp and Gonzalez, Joseph E and Stoica, Ion. Tune: A Research Platform for Distributed Model Selection and Training. *arXiv preprint arXiv:180705118*. 2018;
51. Rudin C, Zhong C. et al. Amazing Things Come From Having Many Good Models. *arXiv preprint arXiv:180705118*. 2024;
52. Müller S, Toborek V, Beckh K, Jakobs M, Bauckhage C, Welke P. An Empirical Evaluation ;Rashomon Effect;Explainable Machine Learning. presented at: Machine Learning and Knowledge Discovery in Databases: Research Track: European Conference, ECML PKDD 2023, Turin, Italy, September 18–22, 2023, Proceedings, Part III; 2023; Turin, Italy. <https://doi.org/10.1007/978-3-031-43418-1_28>
53. Paes LM, Cruz R, Calmon FP, Diaz M. On the Inevitability of the Rashomon Effect. 2023:549-554.
54. Kobyli'nska K, Krzyzi'nski M, Machowicz R, Adamek M, Biecek P. Exploration of the Rashomon Set Assists Trustworthy Explanations for Medical Data. 2023:
55. Yamamoto Y, Tsuzuki T, Akatsuka J, et al. Automated acquisition of explainable knowledge from unannotated histopathology images. *Nature Communications*. 2019/12/18 2019;10(1):5642. doi:10.1038/s41467-019-13647-8
56. Amgad M, Hodge JM, Elsebaie MAT, et al. A population-level digital histologic biomarker for enhanced prognosis of invasive breast cancer. *Nature Medicine*. 2024/01/01 2024;30(1):85-97. doi:10.1038/s41591-023-02643-7
57. D'Amour A, Heller K, Moldovan D, et al. Underspecification presents challenges for credibility in modern machine learning. *J Mach Learn Res*. 2022;23(1):Article 226.

# SUPPLEMENTARY DATA FOR DEEP LEARNING FOR FETAL INFLAMMATORY RESPONSE DIAGNOSIS IN THE UMBILICAL CORDS

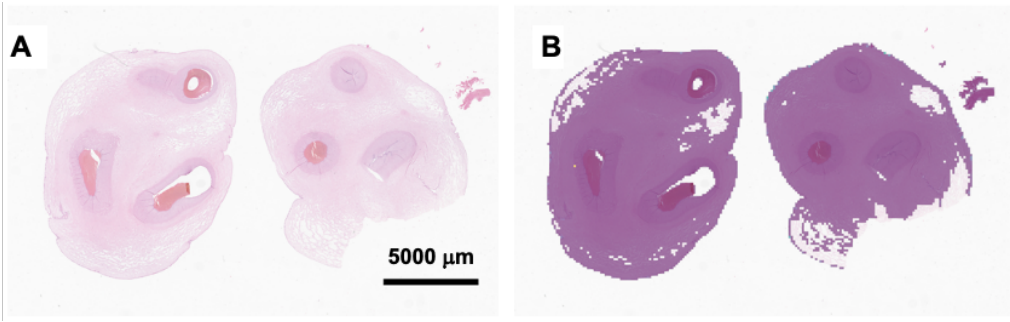

**Supplementary Fig. 1.** Attention Maps of FIR predictions. A: FIR 0 umbilical cord sample. B: The attention map of FIR 0 shows a new-uniform low attention across the tissue.

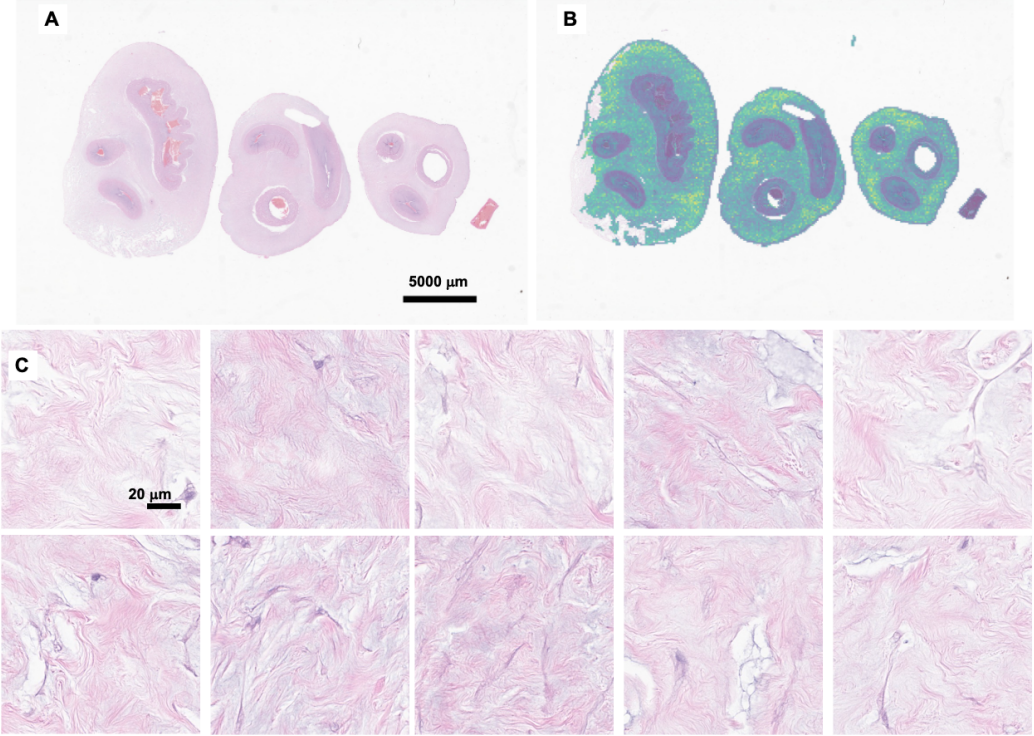

**Supplementary Fig. 2.** Attention Maps of FIR predictions. A: FIR 1 umbilical cord sample. B: The attention map for FIR 1 highlights regions across the stroma of the Wharton's Jelly. C: Top 10 high-attention patches, show a mixture of normal and activated-appearing stroma within the Wharton's jelly.

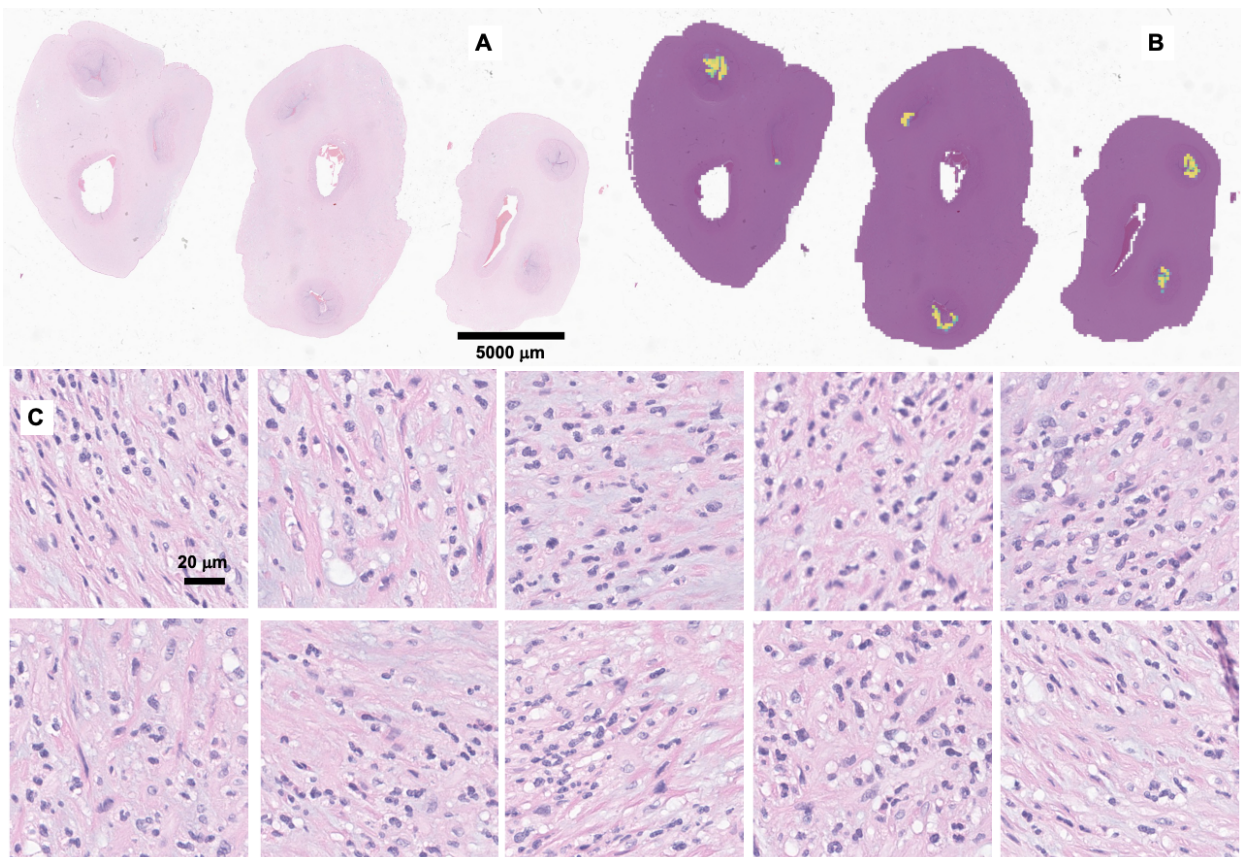

**Supplementary Fig. 3.** Attention Maps of FIR2 predictions. A: FIR 2 sample with umbilical arteritis. B: The attention map shows concentrated focus in the arteries. C: Top 10 high attention patches show neutrophils in the umbilical arteries.

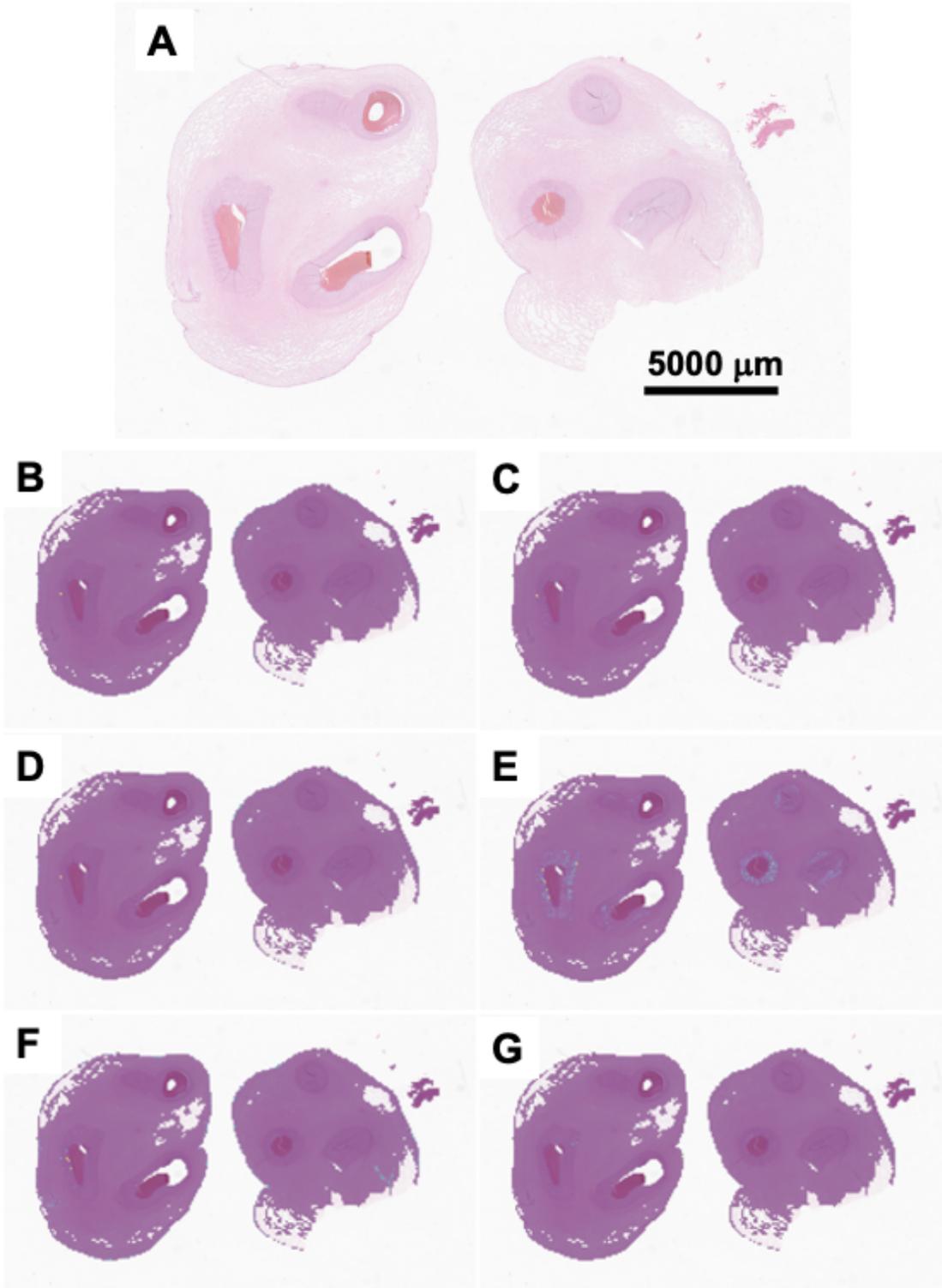

**Supplementary Fig. 4.** FIR 0 Sample with correct predictions across all ensembled models. A: The original histology image. B-G: Attention maps for models ranked from the highest balanced accuracy model to the lowest, as included in the ensemble.

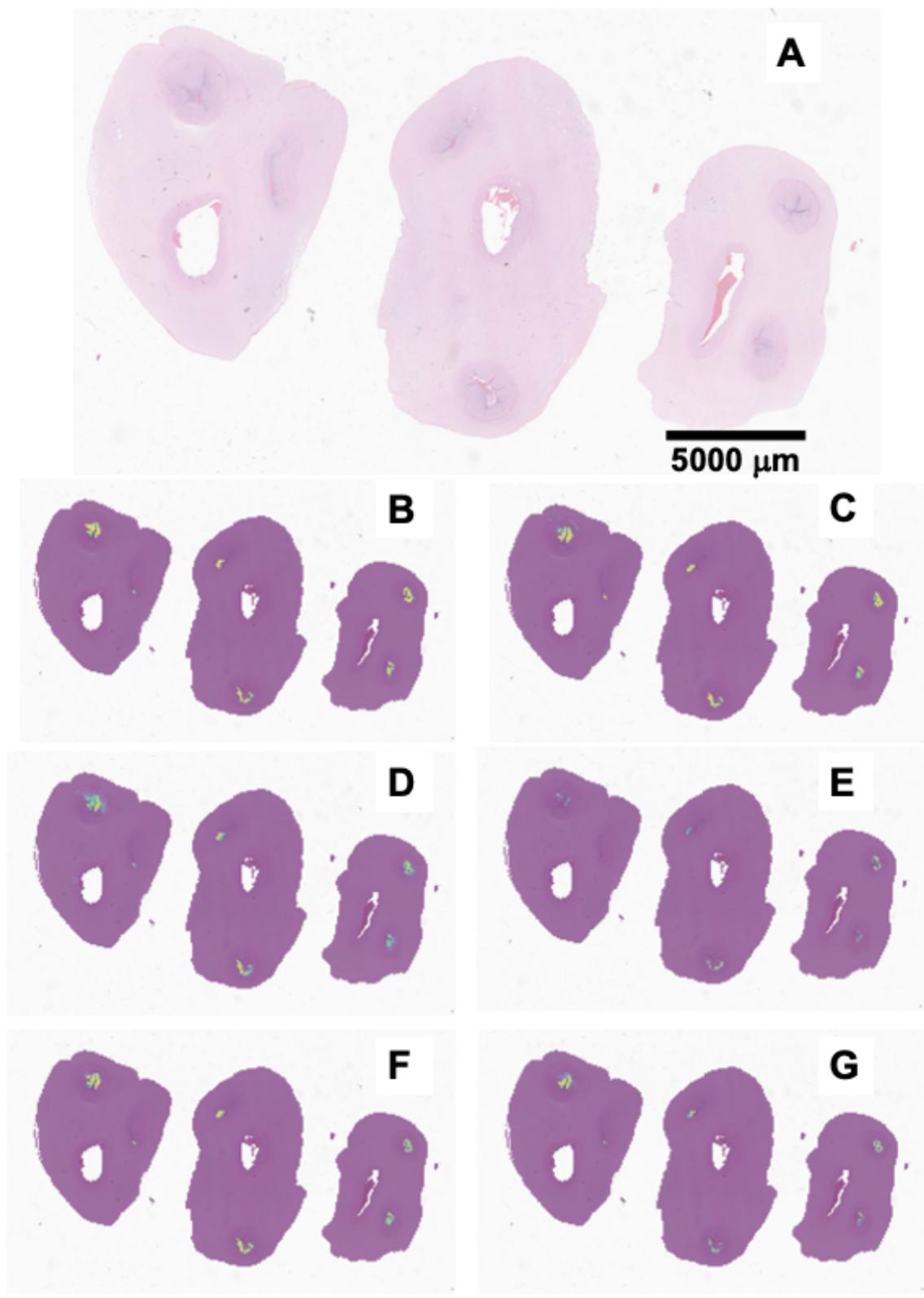

**Supplementary Fig. 5.** FIR 2 Sample with correct predictions across all ensembled models. A: The original histology image. B-G: Attention maps for models ranked from the highest balanced accuracy model to the lowest, as included in the ensemble.